# Strengthening Cybersecurity Resilience in Agriculture Through Educational Interventions: A Case Study of the Ponca Tribe of Nebraska

George Grispos[1], Logan Mears[1], Larry Loucks[2]

[1]University of Nebraska-Omaha, USA

[2] Ponca Economic Development Corporation, USA

ggrispos@unomaha.edu
lmears@unomaha.edu
lloucks@pedco-ne.org

**Abstract:** The increasing digitization of agricultural operations has introduced new cybersecurity challenges for the farming community. This paper introduces an educational intervention called Cybersecurity Improvement Initiative for Agriculture (CIIA), which aims to strengthen cybersecurity awareness and resilience among farmers and food producers. Using a case study that focuses on farmers from the Ponca Tribe of Nebraska, the research evaluates pre- and post- intervention survey data to assess participants' cybersecurity knowledge and awareness before and after exposure to the CIIA. The findings reveal a substantial baseline deficiency in cybersecurity education among participants, however, post-intervention assessments demonstrate improvements in the comprehension of cybersecurity concepts, such as password hygiene, multi-factor authentication, and the necessity of routine data backups. These initial findings highlight the need for a continued and sustained, community-specific cybersecurity education effort to help mitigate emerging cyber threats in the agricultural sector.

## 1. Introduction

The modernization of agriculture through the introduction of digital technologies, known as smart farming and precision agriculture, has revolutionized traditional farming methods. Machinery that once functioned independently now seamlessly integrates with wireless networks, smartphone applications and cloud-based systems (Ayaz et al., 2019). This technological evolution has assisted famers to enhance productivity, streamline resource management, reduce operational costs and waste, as well as automating essential farming activities (Mercier, 2021, Barreto and Amaral, 2018, Meticulous Research, 2022). As a result, industry projections indicate that the Agriculture Internet of Things market is expected to increase from $11.4 billion in 2021, to $18.1 billion in 2026 (Markets and Markets, n.d.).

While smart farming and precision agriculture technologies offer substantial benefits, they also introduce significant cybersecurity risks (Gupta et al., 2020, Sontowski et al., 2020, Angyalos et al., 2021, Freyhof, 2022). Researchers have warned that cybercriminals could exploit vulnerabilities in GPS-guided planting systems, potentially leading to large-scale disruptions in crop production (Freyhof, 2022). More concerning, however, is the growing number of ransomware attacks targeting the digital infrastructure that supports modern farming operations (Doyle, 2022, Bunge, 2021, Plume and Bing, 2021). One potential explanation for this phenomenon is that past food and agriculture policies failed to prioritize cybersecurity threats (Internet Security Alliance, n.d.). For example, both the United States Department of Agriculture (USDA) and the Food and Drug Administration considered cybersecurity a low priority for the sector, only revising this stance in 2015 (Internet Security Alliance, n.d.). However, by then, significant problems had already emerged in the sector. A 2018 survey (Spaulding and Wolf, 2018) of Illinois farmers revealed that many either lacked essential cybersecurity software or were unaware of its presence on their farm systems. Similarly, a Department of Homeland Security report (2018) indicated that farmers and precision agriculture technology providers across the United States (US) did not fully grasp the cyber risks associated with emerging technologies, nor did they take adequate steps to mitigate these threats. As a result, many farms lack essential updates and security measures within their digital infrastructure. The severity of these concerns has been underscored by recent high-profile ransomware incidents (e.g., the Dole Food Company attack (Gatlan, 2023)), which could have been prevented with the implementation of basic cybersecurity practices (Cybersecurity and Infrastructure Security Agency, 2022).

Although cybersecurity challenges affect all agricultural stakeholders, indigenous farmers face unique risks that could severely impact their communities. According to the USDA Census of Agriculture (United States Department of Agriculture, n.d.), there are 5,037 farms owned by American Indians in the US, with some of these located in Knox County, Nebraska. Many of the American Indian farmers in Knox County are members of the Ponca Tribe of Nebraska. Like other American Indian farms in the US, many of the Ponca farms are family-operated and help supply essential food to both the tribe and surrounding community. In fact, the only grocery store in Niobrara, Nebraska, depends on produce from Ponca farmers (Dockendorf, 2020). Given that Knox County is already designated as a "food desert" (Nebraska Legislature, n.d.), a cybersecurity incident affecting these farmers could have serious consequences for local food security. Hence, as cyber threats continue to escalate and technology becomes increasingly embedded in agriculture, it is crucial to equip indigenous food producers (such as Ponca farmers) with cybersecurity knowledge.

This paper presents an educational intervention called Cybersecurity Improvement Initiative for Agriculture (CIIA), which includes developing and implementing a ten-module cybersecurity course specifically designed for American Indian stakeholders, with the goal of improving cybersecurity awareness within the agricultural sector. This study extends ideas initially proposed in the conference publication (Grispos et al., 2025), through a more detailed description of the proposed course, as well as the implementation and evaluation of the course with relevant stakeholders. Relevant information from that conference publication has been included for completeness. The goal of the CIIA is to educate and train individuals that are a part of the American Indian food and agriculture ecosystem on cybersecurity challenges in modern agriculture, with the objective of enhancing their awareness and resilience against cyber threats. Broadly, CIIA seeks to strengthen cybersecurity preparedness with individuals, who have historically had limited access to cybersecurity education programs (Finch et al., 2023, McGill et al., 2022).

The remainder of this paper is structured as follows. Section two describes relevant research related to securing the food and agricultural sector, along with a summary of previous educational efforts in other critical infrastructure domains. Section three details the proposed cybersecurity course through the lens of the CIIA, which includes developing and deploying the course, along with describing evaluation methods to assess the impact of the CIIA. Section four presents the results of the course evaluation, while Section five concludes the work and presents ideas for future research.

## 2. Related Work

As technological advancements continue to transform the agricultural industry, the integration of various digital and automated systems into farming operations introduces significant cybersecurity challenges. As a result, both industry (Department of Homeland Security, 2018) and academia have examined these risks in detail. For example, Demestichas et al. (2020) conducted a survey on technical cybersecurity threats in smart farming, emphasizing the potential risks these threats pose to the agricultural sector. Similarly, Freyhof et al. (2025) examined the vulnerabilities of CAN bus-driven farming equipment, highlighting the financial losses that farmers could suffer if cyberattacks targeting these systems are successful. Gupta et al. (2020) further stressed that Internet of Things devices, which are increasingly embedded in farm operations, are often designed without adequate security measures, making them susceptible to exploitation for malicious activities, including "agri-terrorism". In response to these growing concerns, researchers have proposed various technical solutions, such as the development of cybersecurity frameworks specifically for farming (Chi et al., 2017), the integration of cybersecurity policies to safeguard agricultural operations (Barreto and Amaral, 2018), and the establishment of cybersecurity testbeds to assess and mitigate potential vulnerabilities (Freyhof et al., 2022).

Despite these efforts, previous research indicates that many farmers and food producers remain unaware of the cybersecurity risks associated with adopting new technologies in their agricultural practices (Dehghantanha et al., 2021, Nikander et al., 2020). Spaulding and Wolf (2018) designed and deployed a study to investigate cybersecurity awareness among farmers and reported that only 10% had received prior cybersecurity training, underscoring the widespread lack of education needed to identify and counter cybersecurity threats. Geil et al. (2018) echoed these concerns, emphasizing the necessity for increased computer security education within the agricultural industry, and suggesting that there is an opportunity for organizations to develop specialized cybersecurity training programs tailored to farmers' needs.

Over the years, various cybersecurity educational initiatives have been implemented across different industries, particularly within US critical infrastructure sectors. Waddell et al. (2024) outlined a cybersecurity training program for the healthcare domain, which draws on methodologies from the commercial aviation industry, incorporating dynamic education delivery methods and focused simulations. Chowdhury and Gkioulos (2021) conducted a systematic literature review on cybersecurity training for critical infrastructure and found that most initiatives prioritize hands-on training scenarios and team-based exercises over traditional informational approaches. Kessler and Ramsay (2013) observed that while many colleges and universities in the United States offer cybersecurity programs, these programs often lack direct applicability to homeland security concerns. As a result, they proposed several modifications to cybersecurity curricula to better serve homeland security students. While substantial research has explored cybersecurity training initiatives for various critical infrastructure sectors, there has been little focus on the development and implementation of cybersecurity education programs specifically for the food and agricultural industry.

## 3. Cybersecurity Improvement Initiative for Agriculture (CIIA)

The Cybersecurity Improvement Initiative for Agriculture (CIIA) is an effort aimed at enhancing cybersecurity awareness and preparedness within the agricultural sector. CIIA involves developing a specialized cybersecurity course, delivering the course through both in-person and online formats, and evaluating its effectiveness using a *pre-test-treatment-post-test experiment* (Campbell and Stanley, 1963) to measure participants' knowledge and cybersecurity readiness before and after completing the training.

### 3.1 Proposed Cybersecurity Course

The proposed cybersecurity course consists of ten modules, designed to be accessible through both in-person and online learning formats. Regardless of the delivery approach, each module features open-source readings and classroom instruction-style videos that effectively present the module content and key concepts. To help enhance understanding and real-world application, case studies and relevant news articles are included where appropriate. These additions help provide a practical context for the material being taught. In addition, to ensure that the course participants obtain exposure to cybersecurity content that will help "safeguard(s) and promote(s) America's national security and economic prosperity" (National Institute of Standards and Technology, 2021), the course content was mapped to the NICE Cybersecurity Workforce Framework (National Institute of Standards and Technology, 2016). While the framework does not include a work role associated with the agriculture industry, individuals taking the modules will develop Knowledge Areas (KAs) and Skills Areas (SAs) from the Framework. Table 1 provides an overview of the content of each module, along with the applicable KAs and SAs. Each module is also described in detail below.

| Module Name | Description Summary | Relevant KAs & SAs |
|---|---|---|
| **Introduction to Cybersecurity** | Defines cybersecurity, introduces the CIA triad, and explores cybersecurity risks in agriculture. Case studies: JBS ransomware attack, and Ontario hog farm cyberattack. | K0002, K0004, K0026, K0044 |
| **Computer Networks and the Internet** | Covers farm computer networks, including LANs, WANs, wireless networks, and vulnerabilities in network devices. | K0001, K0011, K0029 |
| **Social Engineering and Other Attacks** | Explores social engineering threats, phishing, identity theft, password attacks, and web-based threats that exploit human psychology. | S0052, K0158 |
| **Malware and Ransomware** | Examines viruses, worms, trojans, ransomware and their impact on farming operations, and recovery strategies. Case study: New Cooperative Inc. ransomware attack. | K0259, K0392, K0480 |
| **Computer Security Software** | Explains the benefits of antivirus solutions, firewalls, and intrusion detection systems to protect farm networks and systems. | S0076, K0049, K0324, K0487 |
| **Mobile Device and Wireless Security** | Covers mobile security risks, Wi-Fi vulnerabilities, and securing smartphones, tablets, and IoT devices on farms. | K0104, K0108, K0137, K0138 |
| **Cyber Warfare** | Examines nation-state cyber threats, hacktivist attacks, and their impact on the agricultural sector. | K0119, K0162 |
| **Responding to Incidents & Problems** | Introduces first responder actions, incident containment, and reporting cybersecurity incidents to authorities. | K0041, K0042 |
| **The Farm and Food Cybersecurity Act** | Discusses proposed cybersecurity regulations, compliance requirements, and risk management in agriculture. | K0003, K0168 |
| **Countermeasures** | Provides best practices for farm cybersecurity, including multi-factor authentication, access controls, and backup strategies. | K0007, K0021, K0210, K0298, K0561 |

**Table 1. Mapping Module Content to NICE Cybersecurity Workforce Framework**

**Module 1 – Introduction to Cybersecurity**. This module defines cybersecurity and introduces the Confidentiality Integrity Availability (CIA) triad, and explores that cybersecurity means in an agricultural context. Individuals explore how digital threats can compromise farm operations, financial transactions, and supply chain logistics. Real-world case studies are used to illustrate the impact of cyberattacks on agriculture, including the JBS ransomware attack in 2021, and a Southwestern Ontario hog farm successfully defending itself against a cyberattack launched by animal-activist hackers.

**Module 2 – Computer Networks and the Internet**. As farming becomes increasingly reliant on smart technology and automated systems, understanding basic networking principles is crucial for securing farm operations. Hence, this module introduces the structure and function of modern computer networks, helping farmers understand how their farm equipment, management systems, and connected devices communicate over the Internet. The module covers Local Area Networks (LANs), Wide Area Networks (WANs), wireless networks, and networking devices, equipping participants with the knowledge to recognize vulnerabilities in their digital infrastructure.

**Module 3 – Social Engineering and Other Attacks**. This module focuses on discussing various types of attacks, including deception-based threats that exploit human psychology rather than technical vulnerabilities. Individuals learn about various social engineering tactics, including phishing, identity theft, and online fraud. The module is also used to introduce password attacks, which target weak or reused passwords to gain unauthorized access to accounts and systems.

**Module 4 – Malware and Ransomware**. This module explores the different type of malware, including viruses, worms, trojans, and ransomware, and examines their potential impact on farming operations. Individuals also learn about how ransomware attacks can disrupt production processes, as well as recovery strategies for ransomware attacks, including incident response planning, backup restoration, and working with cybersecurity professionals to recover encrypted data.

**Module 5 – Computer Security Software**. This module focuses on security tools that can help protect farm systems from various cyber threats, including antivirus programs, firewalls, and intrusion detection systems. This module also attempts to emphasize the importance of regular software updates, patch management, and security monitoring to ensure that farm networks remain resilient against evolving cyber threats.

**Module 6 – Mobile Device and Wireless Security**. As farmers increasingly rely on smartphones, tablets, and wireless-controlled equipment, mobile device and wireless network security becomes a critical area of focus. This module discusses the risks associated with mobile devices, threats from open Wi-Fi networks, and methods for hardening access points to prevent unauthorized access. Individuals also learn about best practices for securing their mobile devices and wireless networks.

**Module 7 – Cyber Warfare**. This module examines nation-state cyber threats, hacktivist attacks, and other advanced cyber threats that could impact agriculture as a critical infrastructure sector. Individuals gain an understanding of the motivations behind cyberattacks and the potential consequences for the food and agricultural sector.

**Module 8 – Responding to Incidents and Problems**. Knowing how to respond to cyber threats is just as important as preventing them, and this module provides individuals with a structured approach to handling security incidents. Participants will learn about first responder actions, incident containment, and recovery strategies, ensuring they are prepared to act quickly in the event of a cyberattack. This module also covers the importance of reporting cyber incidents to authorities and industry partners.

**Module 9 – The Farm and Food Cybersecurity Act**. This module explores proposed legislation designed to enhance the security of the agricultural sector. Hence, individuals will learn about the broader implications of the proposed Farm and Food Cybersecurity Act.

**Module 10 – Countermeasures**. The final module presents a comprehensive overview of best practices and proactive security strategies that individuals can implement to protect their farm operations. Topics include improving employee cybersecurity awareness, developing access control policies, implementing multi-factor authentication, ensuring frequent backups are undertaken to help reduce the impact of an attack.

The learning outcomes from the ten modules are classified into two categories, according to Bloom's taxonomy (Armstrong, 2010): *remember, understand* and, *apply*. For example, after the completion of the module 'Introduction to Cybersecurity', students should *remember and understand* the cybersecurity principles of confidentiality, integrity, availability, while after completing the 'Malware and Ransomware' module, students should be able to *apply* implement basic cyber hygiene practices to minimize the impact of these threats.

## 3.2 Cybersecurity Course Delivery

Individuals accessed the cybersecurity course through either an online workforce development platform or by attending an in-person workshop. The Ponca Economic Development Corporation (PEDCO) offers a variety of workforce development opportunities for the Ponca Tribe, other American Indian tribes, and the public. These educational initiatives are traditionally delivered through a third-party platform, which supports both existing programs and the creation of customized educational offerings. Hence, the proposed cybersecurity course was added to this platform, as an additional opportunity, to educate and train individuals on cybersecurity concerns associated with modern-day farm settings.

PEDCO also offers educational programs in computer hardware, operating systems, and computer networking. Individuals unfamiliar with these foundational topics were encouraged to complete these courses before enrolling in the cybersecurity training to establish a baseline understanding of key technical concepts. This approach ensured participants were better prepared to engage with and apply the cybersecurity material effectively.

To promote the course, PEDCO leveraged existing Tribal communication channels to reach the Ponca Tribe of Nebraska and other American Indian food producers across the state. Interested individuals were given access to the course modules and the corresponding evaluation mechanisms, ensuring a structured learning experience that measured participants' progress and comprehension.

In addition to the online delivery of the cybersecurity course, in-person workshops were organized to provide an alternative learning format. This decision was based on initial invited participant feedback, which indicated a preference for face-to-face instruction over online learning. As a result, interested individuals were invited to attend workshop sessions covering the same module content. Like the online course, workshop participants were also given access to the evaluation mechanisms to help assess their knowledge before and after exposure to the course content. Hence, to maximize outreach, three workshops were conducted between October 2024 and February 2025, which helped to further disseminate the cybersecurity content to a wider audience.

## 3.3 Cybersecurity Course Evaluation

To assess the effectiveness of the cybersecurity course among the course participants, a pre-test-post-test experimental design is implemented (Campbell and Stanley, 1963). Before participating in the course, the course participants complete a pretest survey to establish a baseline measurement of their existing cybersecurity knowledge and awareness. This initial assessment provides insight into their understanding of cybersecurity risks and protective measures before exposure to the cybersecurity course. The participants are then subjected to the *treatment*, the course. Following the completion of the course, participants undergo a post-test survey to evaluate any changes in their cybersecurity knowledge and awareness. This comparative analysis determines whether the course enhances participants' understanding of cybersecurity threats and best practices. All surveys are conducted in compliance with the University of Nebraska's Institutional Review Board (IRB) protocols (IRB# 0675-24), which ensures that ethical considerations are met before any data collection is undertaken with the course participants.

The pre-test and post-test surveys were conducted using a web-based survey hosted on Qualtrics (Oates et al., 2022). The pre-test survey (hereafter referred to as the "Entry Survey") aimed to assess participants' baseline cybersecurity awareness before beginning the course, as well as their prior experience with cybersecurity education and knowledge. The post-test survey (hereafter referred to as the "Exit Survey") evaluated the impact of the cybersecurity course, measuring any changes in participants' understanding of cybersecurity threats and risks as well as cybersecurity best practices in farm settings. Both surveys collected demographic information and underwent a validation process to mitigate potential researcher bias (Pfleeger & Kitchenham, 2001). To ensure clarity and question appropriateness, both survey questions were reviewed by two individuals, one in the cybersecurity domain, and the other in agricultural education. Based on their feedback, minor revisions were made, including simplifying questions and expanding response options for closed-ended questions. Participants completed the Entry Survey before gaining access to the course content, while the Exit Survey was made available upon course completion.

## 4. Cybersecurity Course Impact

A total of twelve individuals volunteered to participate in the Cybersecurity Improvement Initiative for Agriculture (CIIA). While this sample size may be considered small, it is important to consider the broader context. The Ponca Tribe of Nebraska has a total population of 5,344 citizens, with only 1,923 currently residing in Nebraska (Ponca Tribe of Nebraska, n.d.). Moreover, according to the 2022 USDA Census of Agriculture (United States Department of Agriculture, 2024), only 115 out of Nebraska's 44,479 farmers identify as American Indian. The exact number of Ponca farmers, however, remains unknown.

All twelve participants successfully completed *both* the Entry and Exit surveys. The following subsections will provide an overview of the participants' demographic characteristics, followed by a detailed examination of the survey results, including key insights and any notable changes between the Entry and Exit responses.

### 4.1 Demographics

Using data from the 2022 USDA Agricultural Census (United States Department of Agriculture, 2024) as a benchmark, initial survey questions were designed to determine participants' age, gender, and ethnic background. In terms of gender, nine out of twelve participants (75%) identified as male, while three (25%) identified as female. Comparatively, the USDA Agricultural Census reports that the U.S. farming population is 65.6% male and 34.4% female (United States Department of Agriculture, 2024). Regarding age, nearly 60% of participants (7 out of 12) reported being 46 years or older. This aligns closely with industry data, which indicates that the average age of a U.S. farmer is 58, with 60% of farmers aged 55 or older. Although the cybersecurity course was primarily aimed at members of the Ponca Tribe of Nebraska, the ethnic composition of participants varied. Seven out of twelve identified as White/Caucasian, four identified as American Indian, and one individual identified as both American Indian and African American.

### 4.2 Entry Survey Results

The initial Entry Survey questions were used to assess prior cybersecurity education and the participant's perceptions of vulnerability to cyber threats. The results revealed that nine out of twelve participants had never received any form of cybersecurity education, whether formal or informal, before this initiative. This finding supports the assumption that individuals in the food and agricultural sector may be more vulnerable to cyber threats due to a lack of cybersecurity awareness.

When asked, "In your opinion, how vulnerable do you believe farmers are to cybersecurity threats?" most participants expressed concern about the sector's exposure to cyber risks. Six out of twelve participants rated farmers as "very vulnerable," while four participants considered them "somewhat vulnerable." Only two participants expressed a neutral stance, and none indicated that farmers are not vulnerable.

When asked about their familiarity with cybersecurity threats, several notable trends emerged. The analysis revealed that well-known threats, such as computer viruses, hacking, and phishing, had higher levels of familiarity among participants, with a greater proportion selecting "Moderately Familiar" or "Extremely Familiar" (see Figure 1). This suggests that these threats, which have been widely publicized and have significantly impacted both organizations and individuals, are generally well understood.

In contrast, more specialized and complex threats, such as supply chain attacks and insider threats, had a higher proportion of participants indicating lower familiarity levels ("Not Familiar" and "Slightly Familiar"). Given the growing number of ransomware attacks targeting the food and agricultural industry (Grispos and Doctor, 2022), familiarity with ransomware threats was relatively evenly distributed across all levels. This result likely reflects increased awareness due to high-profile incidents in recent years. However, it also underscores the need for greater knowledge and expertise regarding ransomware threats.

The Entry Survey also attempted to identify the types of data that participants believe are most important for American Indian farmers to protect from a cybersecurity perspective, as well as the challenges associated with safeguarding this information. Participants identified several critical data types requiring protection, including:

- Financial and banking records (10 out of 12 participants)
- Personal family information (6 out of 12 participants)
- Tribal farming techniques and knowledge (5 out of 12 participants)
- Yield data or crop management records (5 out of 12 participants)

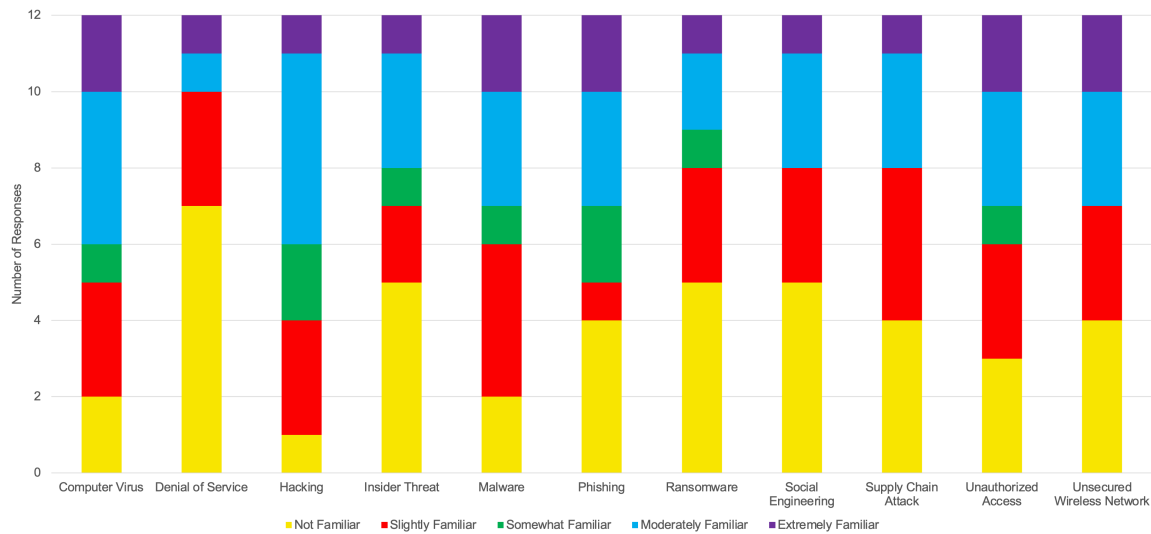

**Figure 1. Familiarity with Cybersecurity Threats**

With participants able to select more than one option, financial and banking records emerged as the most frequently cited asset, indicating concerns about protecting economic information. The emphasis on personal family information (selected by half of the participants) highlights the importance of privacy and security for family-run farms. Additionally, five participants underscored the significance of safeguarding tribal farming techniques and knowledge, reflecting the importance of protecting traditional agricultural practices. These findings suggest that participants are concerned with both conventional cybersecurity risks (e.g., financial and personal data security) and the protection of culturally significant agricultural knowledge (e.g., tribal farming approaches).

Regarding the challenges that American Indian farmers face in implementing cybersecurity measures to protect the information identified in the previous question, several key barriers were identified including:

- Limited knowledge of cybersecurity practices (10 out of 12 participants)
- Cost and affordability of cybersecurity solutions (8 out of 12 participants)
- Lack of time to focus on cybersecurity (6 out of 12 participants)

The most prevalent issue was a lack of knowledge about cybersecurity best practices. This finding aligns with previous studies (Geil et al., 2018, Spaulding and Wolf, 2018) indicating that many farmers lack awareness or training in cybersecurity risk mitigation. Additionally, eight participants expressed concerns about the cost of cybersecurity solutions, suggesting that financial constraints may prevent farmers from investing in security measures or hiring cybersecurity expertise. Another notable challenge, identified by six participants, was the lack of time to focus on cybersecurity. Farming is a time-intensive profession, particularly during planting and harvesting seasons. As a result, these participants indicated that cybersecurity is often deprioritized in favour of more immediate farming responsibilities. The findings suggest that cybersecurity may be perceived as overwhelming for farmers already managing demanding workloads.

The final survey question asked participants whether they believed American Indian farmers face greater cybersecurity risks compared to other farmers. For the six participants who believed that American Indian farmers face greater cybersecurity risks, the primary reasons cited were limited internet connectivity in tribal areas and restricted access to technical support for both general farming operations and cybersecurity needs. These findings underscore the need for further investigation into the specific cybersecurity challenges facing American Indian farmers, particularly regarding infrastructure limitations and access to cybersecurity resources.

### 4.3 Exit Survey Results

The Exit Survey aimed to assess participants' comprehension of the cybersecurity concepts covered in the Cybersecurity Course and evaluate the course's effectiveness in raising awareness of critical cybersecurity issues within the food and agriculture sector. To achieve this, the first question in the survey asked participants whether the course had increased their awareness of cybersecurity threats and potential countermeasures relevant to the industry. Responses were recorded using a five-point Likert scale, ranging from "Strongly Disagree" to "Strongly Agree."

The results, summarized in Figure 2, indicate a positive impact of the course, with all twelve participants selecting either "Strongly Agree" or "Agree" for each cybersecurity issue presented. Specifically, regarding increased awareness of bad actors targeting the food and agriculture sector, seven out of twelve participants responded, "Strongly Agree", while the remaining five selected "Agree." Similarly, increased awareness of specific cybersecurity threats affecting the industry improved, with eight participants strongly agreeing and four agreeing.

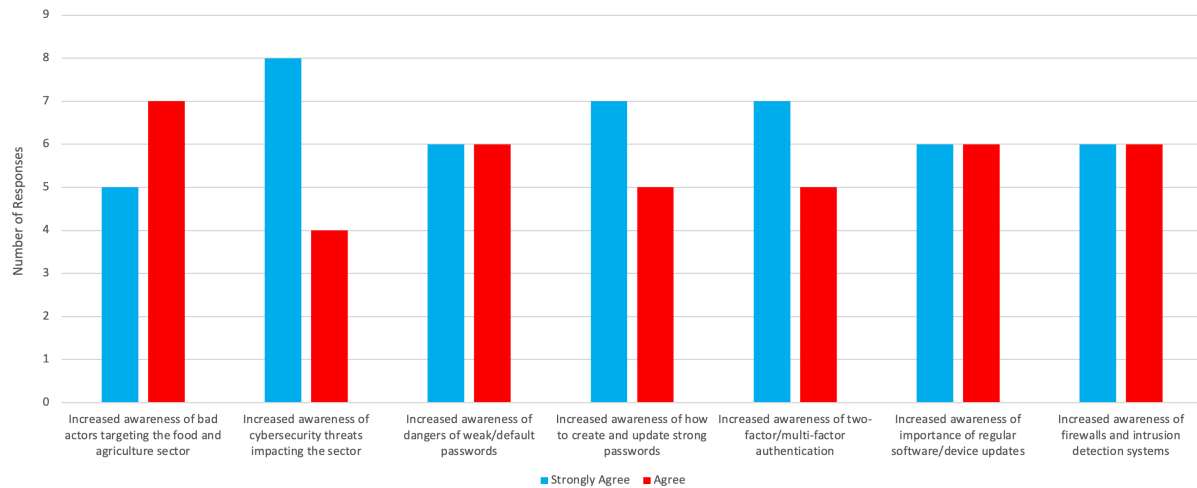

**Figure 2. Increased Familiarity with Cybersecurity Issues**

Beyond general threat awareness, participants also reported an enhanced understanding of password security best practices, including the risks associated with weak or default passwords, the importance of creating strong passwords, and the necessity of regularly updating passwords. Furthermore, the survey results suggest that the course was effective in increasing participants' knowledge of two-factor/multi-factor authentication as an additional layer of security. Another key area of improvement was the participants' increased recognition of the significance of regular software and device updates in mitigating cyber threats. The participants also expressed increased awareness of how firewalls and intrusion detection systems contribute to network security by identifying and preventing unauthorized access.

In addition to assessing participants' awareness of general cybersecurity threats, the survey also evaluated their understanding of two critical security measures: the importance of employee education in cybersecurity and the necessity of creating backups for critical data and applications. Both topics are essential components of a comprehensive cybersecurity strategy and are discussed in the course content. The survey results indicate a significant improvement in participants' awareness of these key practices. When asked about increased awareness regarding the role of employee education in cybersecurity, seven out of twelve participants selected "Strongly Agree", while four participants chose "Agree", demonstrating a widespread increase in awareness regarding the need for continuous training and awareness programs. Only one participant remained "Undecided", suggesting that while the course was effective, some individuals may still require further clarification or reinforcement on this topic.

Similarly, increased awareness regarding the importance of creating backups for safeguarding critical data and applications also improved. Seven participants selected "Strongly Agree", while five participants chose "Agree". Given the increasing number of ransomware attacks targeting agricultural operations and supply chain systems, these findings underscore the value of understanding the importance of regular and secure data backups as a protective measure against data loss, system failures, and security incidents.

Despite the relatively small sample size, these results suggest that the cybersecurity course successfully enhanced participants' understanding of essential security concepts and best practices. By reinforcing the importance of employee training and data backup protocols, the course equipped participants with actionable strategies that can hopefully be implemented within their respective farming operations.

The survey results indicate a high level of satisfaction with the CIIA among participants. When asked about their likelihood of recommending the cybersecurity course to other farmers, seven out of twelve participants selected "Very Likely", while four chose "Somewhat Likely", and only one participant indicated "Somewhat Unlikely". These findings suggest that most participants found the course valuable and would encourage others in the agricultural sector to take it.

Additionally, participants provided verbal feedback and suggestions on ways to improve the course. Several participants emphasized the need for more hands-on elements, with two specifically mentioning that interactive or practical exercises would enhance their learning experience. Others suggested expanding the course to cover mobile device security, as many farmers rely on smartphones and tablets for farm management. One participant expressed interest in learning more about cybersecurity for online activities (such as web browsing etc.), while another highlighted the importance of knowing more of how to respond when something goes wrong in a cybersecurity incident. It appears this participant wanted further details as compared to what was covered in the course content.

## 5. Conclusions and Future Work

This paper presents a Cybersecurity Improvement Initiative for Agriculture (CIIA), which includes developing and implementing a ten-module cybersecurity course specifically designed for American Indian stakeholders, with the goal of improving cybersecurity awareness within the agricultural sector. Using pre-course and post-course surveys, before exposure to the cybersecurity course most participants had no prior cybersecurity education, which suggests that the participants lacked awareness of cybersecurity risks. The post-course survey, however, demonstrated a significant improvement in cybersecurity awareness, with all participants reporting increased familiarity with key threats and best practices. Notably, participants gained a stronger understanding of password security, multi-factor authentication, software updates, and network security measures, suggesting that the course successfully provided them with practical cybersecurity knowledge.

While the course was well received, feedback from participants suggests opportunities for improvement. Several individuals expressed a desire for more hands-on learning, such as interactive exercises or simulations, which could enhance both engagement and, potentially, knowledge retention. Given the high likelihood that participants would recommend the course to others, future work plans to address these concerns and build on this momentum by offering follow-up training and expanding the course offering to the wider agricultural community in the State of Nebraska.

## Acknowledgements

This research was support by the U.S. National Science Foundation (NSF) through the Award #2335812: "Education DCL: EAGER: Enhancing Cybersecurity Awareness of American Indian Farmers and Food Producers: The Ponca Tribe of Nebraska as a Case Study". The statements, opinions, and content included in this publication do not necessarily reflect the position or the policy of the NSF, and no official endorsement should be inferred.

PREPRINTMercier, S. 2021. *Cyber Security Concerns in the U.S. Agricultural Sector*. Available: https://www.agweb.com/opinion/cyber-security-concerns-us-agricultural-sector

Meticulous Research. 2022. *Agriculture IoT Market Report*. Available: https://www.globenewswire.com/news-release/2022/01/20/2369892/0/en/Agriculture-IoT-Market-Worth-22-6-Billion-by-2028-Exclusive-Report-by-Meticulous-Research.html

National Institute of Standards and Technology 2016. National Initiative for Cybersecurity Education Cybersecurity Workforce Framework.

National Institute of Standards and Technology. 2021. *NICE Framework Strategic Plan (2021-2025)*. Available: https://www.nist.gov/itl/applied-cybersecurity/nice/about/strategic-plan

Nebraska Legislature. n.d. *Food Deserts in Nebraska*. Available: https://news.legislature.ne.gov/lrd/files/2015/12/lrd_mow_11.pdf

Nikander, J., Manninen, O. & Laajalahti, M. 2020. Requirements for cybersecurity in agricultural communication networks. *Computers and Electronics in Agriculture,* 179**,** 105776.

Plume, K. & Bing, C. 2021. *Iowa farm services firm: systems offline due to cybersecurity incident*. Available: https://www.reuters.com/technology/iowa-farm-services-company-reports-cybersecurity-incident-2021-09-20/

Ponca Tribe of Nebraska. n.d. *Tribal Enrollment*. Available: https://www.poncatribe-ne.gov/services/tribal-enrollment/ [Accessed].

Sontowski, S., Gupta, M., Chukkapalli, S. S. L., Abdelsalam, M., Mittal, S., Joshi, A. & Sandhu, R. Cyber attacks on smart farming infrastructure.  6th International Conference on Collaboration and Internet Computing, 2020. IEEE, 135-143.

Spaulding, A. D. & Wolf, J. R. 2018. Cyber-Security Knowledge and Training Needs of Beginning Farmers in Illinois. Agricultural and Applied Economics Association.

United States Department of Agriculture 2024. 2022 Census of Agriculture.

United States Department of Agriculture. n.d. *USDA Census of Agriculture Historical Archive*. Available: https://agcensus.library.cornell.edu

Waddell, M. Human factors in cybersecurity: Designing an effective cybersecurity education program for healthcare staff.  Healthcare Management Forum, 2024. SAGE Publications Sage CA: Los Angeles, CA, 13-16.